\documentclass[draft,onecolumn,romanappendices]{IEEEtran}

\usepackage[english]{babel} % Different languages:
\usepackage{amsmath}                % AMS Fonts
\usepackage{amsfonts}               % AMS Fonts 
\usepackage{amssymb}                % Mathemat. symbols of AMS
\usepackage{amsopn}                 % Further math. operators like
\usepackage{dsfont}
\allowdisplaybreaks[3] % Page breaks within eqnarray and IEEEeqnarray
\usepackage{mathrsfs}               % new mathfonts: \mathscr{ABC}
\usepackage{calc}                   % Calculations within the tex-file
\usepackage{graphicx}        % Further graphic-commands (use
\usepackage{verbatim}               % citations directly included
\usepackage{macros}
\usepackage{color}
\usepackage{url}

\newtheorem{theorem}{Theorem}
\newtheorem{lemma}{Lemma}

\renewcommand{\I}[1]{\mathds{1}\!\left\{#1\right\}}

\title{The Shannon Lower Bound\\ is Asymptotically Tight}

\author{Tobias~Koch,~\IEEEmembership{Member,~IEEE}

\thanks{T.~Koch has been supported in part by a Marie Curie Career Integration Grant through the 7th
European Union Framework Programme under Grant 333680, by the Ministerio de Econom\'ia of Spain under Grants TEC2013-41718-R, RYC-2014-16332, and TEC2015-69648-REDC, and by the Comunidad de Madrid under Grant S2013/ICE-2845. This paper will be presented in part at the 2016 International Zurich Seminar on Communications, Zurich, Switzerland.}

\thanks{The author is with the Signal Theory and Communications Department, Universidad Carlos III de Madrid, 28911, Legan\'es, Spain and also with the Gregorio Mara\~n\'on Health Research Institute, Madrid, Spain (e-mail: \texttt{koch@tsc.uc3m.es}).}}

\begin{document}

\maketitle

\begin{abstract}
The \emph{Shannon lower bound} is one of the few lower bounds on the rate-distortion function that holds for a large class of sources. In this paper, it is demonstrated that its gap to the rate-distortion function vanishes as the allowed distortion tends to zero for all sources having a finite differential entropy and whose integer part is finite. Conversely, it is demonstrated that if the integer part of the source has an infinite entropy, then its rate-distortion function is infinite for every finite distortion. Consequently, the Shannon lower bound provides an asymptotically tight bound on the rate-distortion function if, and only if, the integer part of the source has a finite entropy.
\end{abstract}


\section{Introduction}
\label{sec:intro}
Suppose that we wish to quantize a memoryless, $d$-dimensional source with a distortion not larger than $D$. More specifically, suppose a source produces the sequence of independent and identically distributed (i.i.d.), $d$-dimensional, real-valued, random vectors $\{\vect{X}_k,\,k\in\Integers\}$ according to the distribution $P_{\vect{X}}$, and suppose that we employ a vector quantizer that produces a sequence of reconstruction vectors $\{\hat{\vect{X}}_k,\,k\in\Integers\}$ satisfying
\begin{equation}
\label{eq:distortion}
\varlimsup_{n\to\infty} \frac{1}{n} \sum_{k=1}^n \E{\bigl\|\vect{X}_k-\hat{\vect{X}}_k\bigr\|^r} \leq D
\end{equation}
for some norm $\|\cdot\|$ and some $r>0$. (We use $\varlimsup$ to denote the \emph{limit superior} and $\varliminf$ to denote the \emph{limit inferior}.) Rate-distortion theory states that if for every blocklength $n$ and distortion constraint $D$ we quantize the sequence of source vectors $\vect{X}_1,\ldots,\vect{X}_n$ to one of $e^{nR(D)}$ possible sequences of reconstruction vectors $\hat{\vect{X}}_1,\ldots,\hat{\vect{X}}_n$, then the smallest rate $R(D)$ (in nats per source symbol) for which there exists a vector quantizer satisfying \eqref{eq:distortion} is given by \cite{shannon59,berger71}
\begin{equation}
\label{eq:R(D)}
R(D) = \inf_{P_{\hat{\vect{X}}|\vect{X}}\colon \E{\|\vect{X}-\hat{\vect{X}}\|^r}\leq D} I\bigl(\vect{X};\hat{\vect{X}}\bigr)
\end{equation}
where the infimum is over all conditional distributions of $\hat{\vect{X}}$ given $\vect{X}$ for which
\begin{equation}
\label{eq:D}
\E{\bigl\|\vect{X}-\hat{\vect{X}}\bigr\|^r} \leq D
\end{equation}
and where the expectation in \eqref{eq:D} is computed with respect to the joint distribution $P_{\vect{X}}P_{\hat{\vect{X}}|\vect{X}}$. Here and throughout the paper we omit the time indices where they are immaterial. The rate $R(D)$ as a function of $D$ is referred to as the \emph{rate-distortion function}.

Unfortunately, the rate-distortion function is unknown except in a few special cases. It therefore needs to be assessed by means of upper and lower bounds. Arguably, for sources with a finite differential entropy, the most important lower bound is the \emph{Shannon lower bound} \cite{shannon59,berger71}, which for a $d$-dimensional, real-valued source and the distortion constraint \eqref{eq:D} is given by \cite{yamada80}
\begin{equation}
\label{eq:SLB}
R_{\text{SLB}}(D) = h(\vect{X}) + \frac{d}{r}\log\frac{1}{D} - \frac{d}{r}\log\left(\frac{r}{d} \bigl(V_d \Gamma(1+d/r)\bigr)^{r/d}e\right).
\end{equation}
Here $\log(\cdot)$ denotes the natural logarithm, $V_d$ denotes the volume of the $d$-dimensional unit ball $\{\vect{x}\in\Reals^d\colon \|\vect{x}\|\leq 1\}$, and $\Gamma(\cdot)$ denotes the Gamma function. While this lower bound is tight only for some special sources, it converges to the rate-distortion function as the allowed distortion $D$ tends to zero, provided that the source satisfies some regularity conditions; see, e.g., \cite{linkov65}--\nocite{gerrishschultheiss64,binia74}\cite{linderzamir94}. A finite-blocklength refinement of the Shannon lower bound has recently been given by Kostina \cite{kostina15,kostina15_arxiv}.

To the best of our knowledge, the most general proof of the asymptotic tightness of the Shannon lower bound is due to Linder and Zamir \cite{linderzamir94}. While Linder and Zamir considered more general distortion measures, specialized to the norm-based distortion \eqref{eq:D}, they showed the following.

\begin{theorem}[Linder and Zamir \protect{\cite[Cor.~1]{linderzamir94}}]
\label{cor:linderzamir}
Suppose that $\vect{X}$ has a probability density function (pdf) and that $h(\vect{X})$ is finite. Assume further that there exists an $\alpha>0$ such that $\E{\|\vect{X}\|^\alpha}<\infty$. Then the Shannon lower bound is asymptotically tight, i.e.,
\begin{equation}
\lim_{D\downarrow 0} \bigl\{R(D) - R_{\textnormal{SLB}}(D)\bigr\} = 0.
\end{equation}
\end{theorem}
\begin{IEEEproof}
See \cite{linderzamir94}.
\end{IEEEproof}
The theorem's conditions are very mild and satisfied by the most common source distributions. In fact, Theorem~\ref{cor:linderzamir} demonstrates that the Shannon lower bound provides a good approximation of the rate-distortion function for small distortions even if there exists no quantizer with a finite number of codevectors and of finite distortion, i.e., when $\E{\|\vect{X}\|^r}=\infty$. However, the theorem's conditions are more stringent than the ones sometimes encountered in analyses of the rate and distortion redundancies of high-resolution quantizers. This is relevant because the Shannon lower bound is often used as a benchmark against which the performance of such quantizers is measured.

For example, Gish and Pierce \cite{gishpierce68} studied the smallest output entropy that can be achieved via scalar quantization with given expected quadratic distortion, i.e.,
\begin{equation}
\label{eq:scalarquant}
R_{\text{s}}(D) = \inf_{q\colon \E{(X-q(X))^2}\leq D} H\bigl(q(X)\bigr)
\end{equation}
where the infimum is over all deterministic mappings $q(\cdot)$ from the source alphabet $\set{X}$ to some (countable) reconstruction alphabet $\hat{\set{X}}$ satisfying $\E{(X-q(X))^2}\leq D$. For one-dimensional sources that have a pdf satisfying some continuity and decay constraints, they showed that the asymptotic excess rate is given by
\begin{equation}
\label{eq:excess}
\varliminf_{D\downarrow 0} \bigl\{R_{\text{s}}(D)-R(D)\bigr\} = \frac{1}{2}\log\frac{\pi e}{6}.
\end{equation}
They further showed that this excess rate can be achieved by a uniform quantizer, hence the well-known result that ``uniform quantizers are asymptotically optimal as the allowed distortion tends to zero." Since the rate-distortion function $R(D)$ is in general unknown, they showed instead that
\begin{equation}
\varliminf_{D\downarrow 0} \bigl\{R_{\text{s}}(D)-R_{\text{SLB}}(D)\bigr\} = \frac{1}{2}\log\frac{\pi e}{6}.
\end{equation}
This is equivalent to \eqref{eq:excess} whenever the Shannon lower bound is asymptotically tight. A dual formulation of \eqref{eq:excess} was given by Zador \cite{zador66} as the smallest asymptotic excess distortion with respect to the distortion-rate function as the rate tends to infinity. While Zador's original derivation was flawed, a rigorous proof of the same result was given by Gray, Linder, and Li \cite{graylinderli02}. In their work, they consider $d$-dimensional source vectors $\vect{X}$ that have a pdf, whose differential entropy is finite, and that satisfy
\begin{equation}
\label{eq:H(Q1)}
H(\lfloor \vect{X} \rfloor) < \infty.
\end{equation}
Here $\lfloor \vect{a} \rfloor$, $\vect{a}=(a_1,\ldots,a_d)\in\Reals^d$ denotes the $d$-dimensional vector with components $\lfloor a_{1} \rfloor,\ldots,\lfloor a_d \rfloor$, and $\lfloor a\rfloor$, $a\in\Reals$ denotes the integer part of $a$, i.e., the largest integer not larger than $a$. In words, condition \eqref{eq:H(Q1)} demands that quantizing the source with a cubic lattice quantizer of unit-volume cells gives rise to a discrete random vector of finite entropy. This ensures that the quantizer output can be further compressed using a lossless variable-length code of finite expected length. Koch and Vazquez-Vilar \cite{kochvazquez15_arxiv} recently demonstrated that these assumptions are also sufficient to recover Gish and Pierce's result \eqref{eq:excess}.

As we shall argue below, \eqref{eq:H(Q1)} is weaker than the assumption $\E{\|\vect{X}\|^\alpha}<\infty$ required in Theorem~\ref{cor:linderzamir} for the asymptotic tightness of the Shannon lower bound. One may thus wonder whether there are sources for which the performance of high-resolution quantizers can be evaluated but the Shannon lower bound does not constitute a relevant performance benchmark. In this paper, we demonstrate that this is not the case. We show that for sources that have a pdf and whose differential entropy is finite, the Shannon lower bound \eqref{eq:SLB} is asymptotically tight if \eqref{eq:H(Q1)} is satisfied. Conversely, we demonstrate that for sources that do not satisfy \eqref{eq:H(Q1)}, the rate-distortion function is infinite for any finite distortion. Hence, condition \eqref{eq:H(Q1)} is necessary and sufficient for the asymptotic tightness of the Shannon lower bound.

The quantity $H(\lfloor\vect{X}\rfloor)$ in \eqref{eq:H(Q1)} is intimately related with the \emph{R\'enyi information dimension} \cite{renyi59}, defined as
\begin{equation}
d(\vect{X}) \triangleq \lim_{m\to\infty} \frac{H\left(\left\lfloor m\vect{X} \right\rfloor/m\right)}{\log m}, \quad \text{if the limit exists}
\end{equation}
which in turn coincides with the \emph{rate-distortion dimension} introduced by Kawabata and Dembo \cite{kawabatadembo94}; see also \cite{wuverdu10_2}. Generalizing Proposition~1 in \cite{wuverdu10_2} to the vector case, it can be shown that the R\'enyi information dimension is finite if, and only if, \eqref{eq:H(Q1)} is satisfied and that a sufficient condition for finite R\'enyi information dimension is $\E{\log(1+\|\vect{X}\|)}<\infty$, which in turn holds for any source vector for which $\E{\|\vect{X}\|^{\alpha}}<\infty$ for some $\alpha>0$. Thus, \eqref{eq:H(Q1)} is indeed weaker than the assumption that $\E{\|\vect{X}\|^{\alpha}}<\infty$.

It is common to assume that the differential entropy of the source is finite, since otherwise the Shannon lower bound \eqref{eq:SLB} is uninteresting. We next briefly discuss how \eqref{eq:H(Q1)} and the assumption of a finite differential entropy are related. As demonstrated, e.g., in the proof of Theorem~3 in \cite{csiszar61}, a finite $H(\lfloor \vect{X} \rfloor)$ implies that $h(\vect{X})<\infty$. In fact, one can show that if \eqref{eq:H(Q1)} holds and the random vector $\vect{X}$ has a pdf, then $h(\vect{X})\leq H(\lfloor\vect{X}\rfloor)$ \cite[Cor.~1]{stotzbolcskei14_sub}. Conversely, one can find sources for which the differential entropy is finite but $H(\lfloor\vect{X}\rfloor)$ is infinite. For example, consider a one-dimensional source with pdf
\begin{equation}
f_X(x) = \sum_{m=2}^{\infty} p_m m\, \I{m\leq x <m+\frac{1}{m}}, \quad x\in \Reals
\end{equation}
where
\begin{subequations}
\begin{IEEEeqnarray}{lCll}
p_m & = & \frac{1}{\const{K} m \log^2 m}, \quad & m=2,3,\ldots\\
\const{K} & = & \sum_{m=2}^{\infty} \frac{1}{m\log^2 m} \quad &
\end{IEEEeqnarray}
\end{subequations}
and $\I{\cdot}$ denotes the indicator function. It is easy to check that for such a source
\begin{equation}
H(\lfloor X \rfloor) = \sum_{m=2}^{\infty} p_m \log\frac{1}{p_m} = \sum_{m=2}^\infty \frac{\log\const{K}+\log m + 2\log\log m}{\const{K}m\log^2 m} = \infty
\end{equation}
and
\begin{equation}
h(X) = - \int_{\Reals} f_X(x)\log f_X(x) \d x = \sum_{m=2}^{\infty} \frac{\log\const{K}+2\log\log m}{\const{K} m \log^2 m} < \infty.
\end{equation}
(See remark after Theorem~1 in \cite[pp.~197--198]{renyi59}.) Thus, for sources satisfying $h(\vect{X})>-\infty$, a finite $H(\lfloor \vect{X} \rfloor)$ implies a finite differential entropy but not \emph{vice versa}.

\section{Problem Setup and Main Result}
\label{sec:main}
We consider a $d$-dimensional, real-valued source $\vect{X}$ with support $\set{X}\subseteq\Reals^d$ whose distribution is absolutely continuous with respect to the Lebesgue measure, and we denote its pdf by $f_{\vect{X}}$. We further assume that $\vect{x}\mapsto f_{\vect{X}}(\vect{x})\log f_{\vect{X}}(\vect{x})$ is integrable, ensuring that the differential entropy
\begin{equation}
h(\vect{X}) \triangleq - \int_{\set{X}} f_{\vect{X}}(\vect{x}) \log f_{\vect{X}}(\vect{x}) \d\vect{x}
\end{equation}
is well-defined and finite. We have the following result.

\begin{theorem}[Main Result]
\label{thm:main}
Suppose that the $d$-dimensional, real-valued source $\vect{X}$ has a pdf and that $h(\vect{X})$ is finite. If $H(\lfloor\vect{X}\rfloor)<\infty$, then the Shannon lower bound is asymptotically tight, i.e.,
\begin{equation}
\label{eq:thm_main}
\lim_{D\downarrow 0} \bigl\{R(D) - R_{\textnormal{SLB}}(D)\bigr\} = 0.
\end{equation}
Conversely, if $H(\lfloor\vect{X}\rfloor)=\infty$, then $R(D)=\infty$ for every $D>0$.
\end{theorem}
\begin{IEEEproof}
See Section~\ref{sec:proof}.
\end{IEEEproof}
Thus, Theorem~\ref{thm:main} demonstrates that the Shannon lower bound is asymptotically tight if, and only if, $H(\lfloor\vect{X}\rfloor)$ is finite.

In all fairness, we should mention that Linder and Zamir presented conditions for the asymptotic tightness of the Shannon lower bound that are weaker than the ones presented in Theorem~\ref{cor:linderzamir}; see \cite[Th.~1]{linderzamir94}. Specifically, they showed that the Shannon lower bound is asymptotically tight if $\vect{X}$ has a pdf, if $h(\vect{X})$ is finite, and if there exists a function $\delta\colon \Reals^d\to [0,\infty)$ satisfying the following:
\begin{itemize}
\item[(i)] The equations
\begin{subequations}
\begin{IEEEeqnarray}{rCl}
a(D) \int_{\Reals^d} e^{-s(D)\delta(\vect{x})}\d\vect{x} & = & 1 \\
a(D) \int_{\Reals^d} \delta(\vect{x}) e^{-s(D)\delta(\vect{x})} \d\vect{x} & = & D 
\end{IEEEeqnarray}
\end{subequations}
have a unique pair of solutions $\bigl(a(D),s(D)\bigr)$ for all $D>0$. Moreover, $a(D)$ and $s(D)$ are continuous functions of $D$.
\item[(ii)] Let $\vect{W}_D$ be a random vector with pdf $\vect{x}\mapsto a(D)e^{-s(D)\delta(\vect{x})}$. Then $\vect{W}_D\Rightarrow \vect{0}$ as $D\to 0$, where we use ``$\Rightarrow$" to denote convergence in distribution and $\vect{0}$ denotes the all-zero vector.
\item[(iii)] Let $\vect{Z}_D$ be a random vector that is independent of $\vect{X}$ and that has the pdf
\begin{equation}
\label{eq:fZ}
f_{\vect{Z}_D}(\vect{z}) = \left(\frac{d}{r}\right)^{\frac{d}{r}-1}\frac{1}{V_d\Gamma(d/r)D^{\frac{d}{r}}} e^{-\frac{d}{rD}\|\vect{z}\|^r}, \quad \vect{z}\in\Reals^d.
\end{equation}
Then $\delta(\cdot)$ satisfies $0<\E{\delta(\vect{X})}<\infty$ and $\E{\delta(\vect{X}+\vect{Z}_D)}$ tends to $\E{\delta(\vect{X})}$ as $D$ tends to zero.
\end{itemize}

It is unclear whether there exists a function $\delta(\cdot)$ with the above properties that allows us to prove the asymptotic tightness of the Shannon lower bound for all source vectors $\vect{X}$ satisfying $H(\lfloor \vect{X} \rfloor) < \infty$ and $|h(\vect{X})|<\infty$. In fact, even if there existed such a function, proving that it satisfies the required conditions may be complicated. Fortunately, the existence of such a function is not essential. Indeed, the proof of Theorem~\ref{thm:main} follows closely the proof of Theorem~1 in \cite{linderzamir94} but avoids the use of $\delta(\cdot)$.

\section{Proof of Theorem~\ref{thm:main}}
\label{sec:proof}
The proof consists of two parts. In the first part, we show that if $H(\lfloor\vect{X}\rfloor)<\infty$, then the Shannon lower bound is asymptotically tight (Section~\ref{sub:direct}). In the second part, we show that if $H(\lfloor\vect{X}\rfloor)=\infty$, then $R(D)=\infty$ for every $D>0$ (Section~\ref{sub:converse}).

\subsection{Asymptotic Tightness}
\label{sub:direct}
In this section, we demonstrate the asymptotic tightness of the Shannon lower bound $R_{\text{SLB}}(D)$ for sources that satisfy $H(\lfloor\vect{X}\rfloor)<\infty$ and $|h(\vect{X})|<\infty$. The first steps in our proof are identical to the ones in the proof of Theorem~1 in \cite{linderzamir94}. To keep this paper self-contained, we reproduce all the steps.

To prove asymptotic tightness of $R_{\text{SLB}}(D)$, we derive an upper bound on $R(D)$ whose gap to $R_{\text{SLB}}(D)$ vanishes as $D$ tends to zero. In view of \eqref{eq:R(D)}, an upper bound on $R(D)$ follows by choosing $\hat{\vect{X}}=\vect{X}+\vect{Z}_D$, where $\vect{Z}_D$ is a $d$-dimensional, real-valued, random vector that is independent of $\vect{X}$ and has pdf \eqref{eq:fZ}. It can be shown that $\vect{Z}_D$ satisfies $\E{\|\vect{Z}_D\|^r}=D$; see, e.g., \cite[Sec.~VI]{yamada80}. It follows that
\begin{IEEEeqnarray}{lCl}
R(D) & \leq & I(\vect{X};\vect{X}+\vect{Z}_D) \nonumber\\
& = & h(\vect{X}+\vect{Z}_D) - h(\vect{Z}_D). \label{eq:ACH_1}
\end{IEEEeqnarray}
Furthermore, by evaluating $h(\vect{Z}_D)$ and comparing the result with \eqref{eq:SLB}, we have
\begin{equation}
\label{eq:ACH_2}
R_{\text{SLB}}(D) = h(\vect{X}) - h(\vect{Z}_D).
\end{equation}
Combining \eqref{eq:ACH_1} and \eqref{eq:ACH_2} gives
\begin{equation}
0 \leq R(D) - R_{\textnormal{SLB}}(D) \leq h(\vect{X}+\vect{Z}_D) - h(\vect{X}).
\end{equation}
Thus, asymptotic tightness of the Shannon lower bound follows by proving that
\begin{equation}
\label{eq:ach_1}
\varlimsup_{D\downarrow 0} h(\vect{X} + \vect{Z}_D) \leq h(\vect{X}).
\end{equation}
To this end, we follow the steps (17)--(21) in \cite{linderzamir94} but with $Y_{\Delta(D)}$ and $Y_{\Delta(0)}$ there replaced by the random vectors $\vect{Y}_D$ and $\vect{Y}_0$ having the respective pdfs
\begin{subequations}
\begin{IEEEeqnarray}{lCl}
f_{\vect{Y}_D}(\vect{y}) & = & \sum_{\vect{i}\in\Integers^d} \Prob\bigl(\lfloor\vect{X}+\vect{Z}_D\rfloor = \vect{i}\bigr) \I{\lfloor\vect{y}\rfloor=\vect{i}}, \quad \vect{y}\in\Reals^d \label{eq:ach_fYD}\\
f_{\vect{Y}_0}(\vect{y}) & = & \sum_{\vect{i}\in\Integers^d} \Prob\bigl(\lfloor\vect{X}\rfloor=\vect{i}\bigr) \I{\lfloor\vect{y}\rfloor=\vect{i}}, \quad \vect{y}\in\Reals^d. \label{eq:ach_fY0}
\end{IEEEeqnarray}
\end{subequations}
It follows that
\begin{equation}
\label{eq:ach_D1}
D(f_{\vect{X}+\vect{Z}_D}\| f_{\vect{Y}_D}) = H(\lfloor \vect{X}+\vect{Z}_D\rfloor) - h(\vect{X}+\vect{Z}_D)
\end{equation}
and
\begin{equation}
\label{eq:ach_D2}
D(f_{\vect{X}}\| f_{\vect{Y}_0}) = H(\lfloor \vect{X}\rfloor) - h(\vect{X})
\end{equation}
where $D(f\|g)$ denotes the relative entropy between the pdfs $f$ and $g$ \cite[Eq.~(9.46)]{coverthomas91}. The random vector $\vect{Z}_D$ has the same pdf as $D^{1/r} \vect{Z}_1$, where $\vect{Z}_1$ denotes $\vect{Z}_D$ for $D=1$. Consequently, $\vect{Z}_D\to\vect{0}$ almost surely as $D$ tends to zero and, hence, also in distribution. Since $\vect{X}$ and $\vect{Z}_D$ are independent, it follows that $\vect{X}+\vect{Z}_D \Rightarrow \vect{X}$ as $D$ tends to zero. Furthermore, since the distribution of $\vect{X}$ is absolutely continuous with respect to the Lebesgue measure and the set $\Integers^d$ is countable, the probability $\Prob(\vect{X}\in\Integers^d)$ is zero, so \cite[Th.~2.8.1, p.~122]{AsDo00}
\begin{equation}
\label{eq:ach_1.5}
\lim_{D\downarrow 0} \Prob\bigl(\lfloor\vect{X}+\vect{Z}_D\rfloor=\vect{i}\bigr) = \Prob\bigl(\lfloor\vect{X}\rfloor=\vect{i}\bigr), \quad \vect{i}\in\Reals^d.
\end{equation}
We thus conclude that $f_{\vect{Y}_D}$ converges pointwise to $f_{\vect{Y}_0}$, which by Scheffe's lemma \cite[Th.~16.12]{billingsley95} implies that $\vect{Y}_D \Rightarrow \vect{Y}_0$ as $D$ tends to zero.

By the lower semicontinuity of relative entropy (see, e.g., the proof of Lemma~4 in \cite{csiszar92} and references therein), it follows that
\begin{equation}
\label{eq:ach_2}
\varliminf_{D\downarrow 0} D(f_{\vect{X}+\vect{Z}_D}\| f_{\vect{Y}_D}) \geq D(f_{\vect{X}}\| f_{\vect{Y}_0}).
\end{equation}
Together with \eqref{eq:ach_D1} and \eqref{eq:ach_D2}, this yields
\begin{equation}
\label{eq:ach_dobi}
\varliminf_{D\downarrow 0} \bigl\{H(\lfloor \vect{X}+\vect{Z}_D\rfloor) - h(\vect{X}+\vect{Z}_D)\bigr\} \geq H(\lfloor \vect{X}\rfloor) - h(\vect{X}).
\end{equation}
Since $H(\lfloor\vect{X}\rfloor)<\infty$ and  $|h(\vect{X})|<\infty$, the claim \eqref{eq:ach_1} follows from \eqref{eq:ach_dobi} by showing that $H(\lfloor\vect{X}+\vect{Z}_D\rfloor)$ tends to $H(\lfloor\vect{X}\rfloor)$ as $D$ tends to zero. To this end, we need the following lemma, which we state in its most general form since it may be of independent interest.

\begin{lemma}
\label{lemma:cont}
Let $\vect{X}$ and $\vect{Z}$ be independent $d$-dimensional random vectors. Assume that $\E{\|\vect{Z}\|^r}<\infty$.
\begin{itemize}
\item[(i)] If $H(\lfloor\vect{X}\rfloor)=\infty$, then $H(\lfloor\vect{X}+\eps\vect{Z}\rfloor)=\infty$  for every $\eps>0$.
\item[(ii)] If $H(\lfloor\vect{X}\rfloor)<\infty$ and $\Prob(\vect{X}\in\Integers^d)=0$, then
\begin{equation}
\lim_{\eps\downarrow 0} H(\lfloor\vect{X}+\eps\vect{Z}\rfloor) = H(\lfloor\vect{X}\rfloor).
\end{equation}
\end{itemize}
\end{lemma}
\begin{IEEEproof}
See appendix.
\end{IEEEproof}

The random vector $\vect{Z}_D$ is independent of $\vect{X}$ and has the same pdf as $D^{1/r}\vect{Z}_1$, where $\vect{Z}_1$ satisfies $\E{\|\vect{Z}_1\|^r}=1$. Furthermore, by assumption, $H(\lfloor\vect{X}\rfloor)<\infty$ and $\Prob(\vect{X}\in\Integers^d)=0$ (since $\vect{X}$ has a pdf and $\Integers^d$ is countable). It thus follows from Part~(ii) of Lemma~\ref{lemma:cont} that
\begin{equation}
\label{eq:newlemma}
\lim_{D\downarrow 0} H(\lfloor\vect{X}+\vect{Z}_D\rfloor) = \lim_{D\downarrow 0} H(\lfloor\vect{X}+D^{1/r}\vect{Z}_1\rfloor) = H(\lfloor\vect{X}\rfloor).
\end{equation}
Combining \eqref{eq:newlemma} with \eqref{eq:ach_dobi} yields \eqref{eq:ach_1}, which in turn demonstrates that the Shannon lower bound is asymptotically tight if $H(\lfloor\vect{X}\rfloor)<\infty$ and $|h(\vect{X})|<\infty$. This proves the first part of Theorem~\ref{thm:main}.

\subsection{Infinite Rate-Distortion Function}
\label{sub:converse}
To prove that $H(\lfloor\vect{X}\rfloor)=\infty$ implies $R(D)=\infty$ for every $D>0$, we show that $I(\vect{X};\hat{\vect{X}})=\infty$ for every pair of random vectors $(\vect{X},\hat{\vect{X}})$ satisfying \eqref{eq:D} and $H(\lfloor\vect{X}\rfloor)=\infty$. To this end, we follow along the lines of the proof of Theorem~6 in \cite[App.~A]{stotzbolcskei14_sub}. Indeed, it follows from the data processing inequality \cite[Cor.~7.16]{gray11} that for any arbitrary $\Upsilon>0$
\begin{equation}
\label{eq:con_1}
I(\vect{X};\hat{\vect{X}}) \geq I\bigl(g_{\Upsilon}(\lfloor\vect{X}\rfloor); \lfloor\hat{\vect{X}}\rfloor\bigr)
\end{equation}
where the function $g_{\Upsilon}\colon \Reals^d \to [-\Upsilon,\Upsilon]^d$ clips its argument to the hypercube $[-\Upsilon,\Upsilon]^d$, i.e.,
\begin{equation}
\label{eq:gUps}
g_{\Upsilon}(\vect{x}) \triangleq \max\{\min\{\vect{x},\mathbf{\Upsilon}\},-\mathbf{\Upsilon}\}, \quad \vect{x}\in \Reals^d.
\end{equation}
In \eqref{eq:gUps}, $\mathbf{\Upsilon}$ denotes the $d$-dimensional vector $(\Upsilon,\ldots,\Upsilon)$, and $\max\{\cdot,\cdot\}$ and $\min\{\cdot,\cdot\}$ denote the component-wise maximum and minimum, respectively. Since $H\bigl(g_{\Upsilon}(\lfloor \vect{X}\rfloor)\bigr)$ is finite, the mutual information on the right-hand side (RHS) of \eqref{eq:con_1} can be written in the form
\begin{equation}
\label{eq:con_2}
I\bigl(g_{\Upsilon}(\lfloor \vect{X}\rfloor); \lfloor\hat{\vect{X}}\rfloor\bigr) = H\bigl(g_{\Upsilon}(\lfloor \vect{X}\rfloor)\bigr) - H\bigl(g_{\Upsilon}(\lfloor \vect{X}\rfloor)\bigm| \lfloor\hat{\vect{X}}\rfloor\bigr)
\end{equation}
which is well-defined.

We first show that the second entropy on the RHS of \eqref{eq:con_2} is bounded in $\Upsilon$ for all pairs of vectors $(\vect{X},\hat{\vect{X}})$ satisfying \eqref{eq:D}. Using basic properties of entropy  together with the fact that the entropy of a function of a random variable is less than or equal to the entropy of the random variable itself \cite[Ex.~5, p.~43]{coverthomas91}, we obtain
\begin{IEEEeqnarray}{lCl}
H\bigl(g_{\Upsilon}(\lfloor \vect{X}\rfloor)\bigm| \lfloor\hat{\vect{X}}\rfloor\bigr) & \leq & H\bigl(\lfloor \vect{X}\rfloor\bigm| \lfloor\hat{\vect{X}}\rfloor\bigr) \nonumber\\
& \leq & H\bigl(\lfloor \vect{X}-\hat{\vect{X}}\rfloor\bigr) + H\bigl(\lfloor \vect{X}\rfloor\bigm|\lfloor\hat{\vect{X}}\rfloor,\lfloor\vect{X}-\hat{\vect{X}}\rfloor\bigr).\label{eq:con_2.5}
\end{IEEEeqnarray}
Since $\E{\log(1+\|\vect{X}-\hat{\vect{X}}\|)}<\infty$ for all $(\vect{X},\hat{\vect{X}})$ satisfying \eqref{eq:D}, generalizing Proposition~1 in \cite{wuverdu10_2} to the vector case yields that
\begin{equation}
\label{eq:con_3}
H\bigl(\lfloor \vect{X}-\hat{\vect{X}}\rfloor\bigr) < \infty.
\end{equation}
Furthermore, denoting $\vect{Y}=\vect{X}-\hat{\vect{X}}$, we obtain
\begin{IEEEeqnarray}{lCl}
H\bigl(\lfloor \vect{X}\rfloor\bigm|\lfloor\hat{\vect{X}}\rfloor,\lfloor\vect{X}-\hat{\vect{X}}\rfloor\bigr) & = & H\bigl(\lfloor \hat{\vect{X}}+\vect{Y} \rfloor\bigm|\lfloor\hat{\vect{X}}\rfloor,\lfloor\vect{Y}\rfloor\bigr) \nonumber\\
& \leq & d \log 2 \label{eq:con_4}
\end{IEEEeqnarray}
since, conditioned on $\lfloor\hat{\vect{X}}\rfloor$ and $\lfloor\vect{Y}\rfloor$, each component of $\lfloor \hat{\vect{X}}+\vect{Y} \rfloor$ can only take on the values $\lfloor \hat{X}_{\ell}\rfloor  + \lfloor Y_{\ell} \rfloor$ and $\lfloor \hat{X}_{\ell}\rfloor  + \lfloor Y_{\ell} \rfloor + 1$ (see also the proof of Proposition~8 in \cite{stotzbolcskei14_sub}). Combining \eqref{eq:con_2.5}--\eqref{eq:con_4} yields
\begin{equation}
\label{eq:con_5}
\sup_{\Upsilon>0}H\bigl(g_{\Upsilon}(\lfloor \vect{X}\rfloor)\bigm| \lfloor\hat{\vect{X}}\rfloor\bigr)  < \infty.
\end{equation}

We next show that if $H(\lfloor\vect{X}\rfloor)=\infty$, then
\begin{equation}
\label{eq:inf_Ups}
\lim_{\Upsilon\to\infty} H\bigl(g_{\Upsilon}(\lfloor \vect{X}\rfloor)\bigr) = \infty.
\end{equation}
Since $\Upsilon>0$ is arbitrary, it then follows from \eqref{eq:con_1} and \eqref{eq:con_2} that
\begin{equation}
I(\vect{X};\hat{\vect{X}}) \geq \varlimsup_{\Upsilon\to\infty} \left\{H\bigl(g_{\Upsilon}(\lfloor \vect{X}\rfloor)\bigr) - H\bigl(g_{\Upsilon}(\lfloor \vect{X}\rfloor)\bigm|\lfloor\hat{\vect{X}}\rfloor\bigr)\right\}
\end{equation}
which by \eqref{eq:con_5} and \eqref{eq:inf_Ups} is infinite. Hence, $I(\vect{X};\hat{\vect{X}})=\infty$ for every pair of random vectors $(\vect{X},\hat{\vect{X}})$ satisfying \eqref{eq:D} and $H(\lfloor\vect{X}\rfloor)=\infty$, which implies that the rate-distortion function $R(D)$ is infinite for every $D>0$.

To prove \eqref{eq:inf_Ups}, we note that
\begin{IEEEeqnarray}{lCl}
H\bigl(g_{\Upsilon}(\lfloor \vect{X}\rfloor)\bigr) & \geq & \sum_{\vect{i}\in\Integers^d} \Prob\bigl(\lfloor\vect{X}\rfloor = \vect{i}\bigr) \log\frac{1}{\Prob\bigl(\lfloor\vect{X}\rfloor = \vect{i}\bigr)} \I{\vect{i}\in (-\Upsilon,\Upsilon)^d}
\end{IEEEeqnarray}
since $\Prob(g_{\Upsilon}(\lfloor\vect{X}\rfloor) = \vect{i}) \log\bigl(1/\Prob(g_{\Upsilon}(\lfloor\vect{X}\rfloor) = \vect{i})\bigr)\geq 0$ for $\vect{i}\notin (-\Upsilon,\Upsilon)^d$ and $\Prob(g_{\Upsilon}(\lfloor\vect{X}\rfloor) = \vect{i})=\Prob(\lfloor\vect{X}\rfloor = \vect{i})$ for $\vect{i}\in(-\Upsilon,\Upsilon)^d$. The claim thus follows from Fatou's lemma \cite[Th.~1.6.8, p.~50]{AsDo00} and because $\I{\vect{i}\in (-\Upsilon,\Upsilon)^d}$ converges pointwise to $\I{\vect{i}\in\Integers^d}$ as $\Upsilon\to\infty$:
\begin{equation}
\varliminf_{\Upsilon\to\infty} H\bigl(g_{\Upsilon}(\lfloor \vect{X}\rfloor)\bigr) \geq H(\lfloor \vect{X}\rfloor) = \infty.
\end{equation}
This proves the second part of Theorem~\ref{thm:main}.

\section{Conclusions}
The Shannon lower bound is one of the few lower bounds on the rate-distortion function that hold for a large class of sources. We have demonstrated that this lower bound is asymptotically tight as the allowed distortion vanishes for all sources having a finite differential entropy and a finite R\'enyi information dimension. Conversely, we have demonstrated that if the source has an infinite R\'enyi information dimension, then the rate-distortion function is infinite for any finite distortion.

Assuming a finite R\'enyi information dimension is tantamount to assuming that quantizing the source with a cubic lattice quantizer of unit-volume cells gives rise to a discrete random vector of finite entropy. The latter assumption is natural in rate-distortion theory and often encountered. To this effect, we have demonstrated that this assumption is not only natural, but it is also a necessary and sufficient condition for the asymptotic tightness of the Shannon lower bound.

For ease of exposition, we have only considered norm-based difference distortion measures, which is less general than the distortion measures studied, e.g., by Linder and Zamir \cite{linderzamir94}. While our analysis could be generalized to more general distortion measures, we have refrained from doing so, because we believe that it would obscure the analysis without offering much more insight.

\appendix

\section{Proof of Lemma~\ref{lemma:cont}}
\label{app:lemma_cont}

\subsection{Proof of Lemma~\ref{lemma:cont}: Part~(i)}

We shall show by contradiction that if $H(\lfloor\vect{X}\rfloor)=\infty$, then $H(\lfloor\vect{X}+\eps\vect{Z}\rfloor)=\infty$ for every $\eps>0$. So let us assume that $H(\lfloor\vect{X}\rfloor)=\infty$ but that there exists an $\eps>0$ such that $H(\lfloor\vect{X}+\eps\vect{Z}\rfloor)<\infty$. It then follows that, for any arbitrary $\Upsilon>0$, the difference $H(\lfloor\vect{X}+\eps\vect{Z}\rfloor)-H\bigl(\lfloor\vect{X}+\eps\vect{Z}\rfloor\bigm|g_{\Upsilon}(\lfloor\vect{X}\rfloor)\bigr)$ is well-defined and equal to $I\bigl(\lfloor\vect{X}+\eps\vect{Z}\rfloor; g_{\Upsilon}(\lfloor\vect{X}\rfloor)\bigr)$. (The function $g_{\Upsilon}(\cdot)$ has been defined in \eqref{eq:gUps}.) Consequently, by the nonnegativity of entropy,
\begin{equation}
\label{eq:app(i)_1}
H(\lfloor\vect{X}+\eps\vect{Z}\rfloor) \geq I\bigl(\lfloor\vect{X}+\eps\vect{Z}\rfloor; g_{\Upsilon}(\lfloor\vect{X}\rfloor)\bigr).
\end{equation}
Furthermore, $H\bigl(g_{\Upsilon}(\lfloor\vect{X}\rfloor)\bigr)$ is finite, so the mutual information on the RHS of \eqref{eq:app(i)_1} can also be written as
\begin{equation}
\label{eq:app(i)_2}
I\bigl(\lfloor\vect{X}+\eps\vect{Z}\rfloor; g_{\Upsilon}(\lfloor\vect{X}\rfloor)\bigr) = H\bigl(g_{\Upsilon}(\lfloor\vect{X}\rfloor)\bigr) - H\bigl(g_{\Upsilon}(\lfloor\vect{X}\rfloor)\bigm| \lfloor\vect{X}+\eps\vect{Z}\rfloor\bigr).
\end{equation}
We next show that
\begin{equation}
\label{eq:app(i)_Upsi}
\sup_{\Upsilon>0} H\bigl(g_{\Upsilon}(\lfloor\vect{X}\rfloor)\bigm| \lfloor\vect{X}+\eps\vect{Z}\rfloor\bigr)<\infty.
\end{equation}
To this end, we follow the steps \eqref{eq:con_2.5}--\eqref{eq:con_4} in Section~\ref{sub:converse}. Indeed, as in \eqref{eq:con_2.5}, it can be shown that
\begin{equation}
\label{eq:app(i)_3}
H\bigl(g_{\Upsilon}(\lfloor\vect{X}\rfloor)\bigm| \lfloor\vect{X}+\eps\vect{Z}\rfloor\bigr) \leq H\bigl(\lfloor\eps\vect{Z}\rfloor\bigr) + H\bigl(\lfloor\vect{X}\rfloor\bigm|\lfloor\vect{X}+\eps\vect{Z}\rfloor,\lfloor\eps\vect{Z}\rfloor\bigr).
\end{equation}
Generalizing Proposition~1 in \cite{wuverdu10_2} to the vector case then yields that the first entropy on the RHS of \eqref{eq:app(i)_3} is finite, since the lemma's assumption $\E{\|\vect{Z}\|^r}<\infty$ implies that $\E{\log(1+\|\eps\vect{Z}\|)}<\infty$. Moreover, following the steps in \eqref{eq:con_4}, the second entropy on the RHS of \eqref{eq:app(i)_3} can be upper-bounded by
\begin{equation}
H\bigl(\lfloor\vect{X}\rfloor\bigm|\lfloor\vect{X}+\eps\vect{Z}\rfloor,\lfloor\eps\vect{Z}\rfloor\bigr) \leq d\log 2.
\end{equation}
The claim \eqref{eq:app(i)_Upsi} thus follows.

Since $\Upsilon>0$ is arbitrary, \eqref{eq:app(i)_1} and \eqref{eq:app(i)_2} give
\begin{equation}
\label{eq:app(i)_4}
H(\lfloor\vect{X}+\eps\vect{Z}\rfloor) \geq \varlimsup_{\Upsilon\to\infty} \left\{ H\bigl(g_{\Upsilon}(\lfloor\vect{X}\rfloor)\bigr) -  H\bigl(g_{\Upsilon}(\lfloor\vect{X}\rfloor)\bigm| \lfloor\vect{X}+\eps\vect{Z}\rfloor\bigr)\right\}.
\end{equation}
However, if $H(\lfloor\vect{X}\rfloor)=\infty$ then, by \eqref{eq:inf_Ups} and \eqref{eq:app(i)_Upsi}, the RHS of \eqref{eq:app(i)_4} is infinite, which contradicts the assumption that there exists an $\eps>0$ such that $H(\lfloor\vect{X}+\eps\vect{Z}\rfloor)<\infty$. This proves Part~(i) of Lemma~\ref{lemma:cont}.

\subsection{Proof of Lemma~\ref{lemma:cont}: Part~(ii)}

Using basic properties of entropy, we obtain
\begin{IEEEeqnarray}{lCl}
H(\lfloor \vect{X}+\eps\vect{Z}\rfloor) & \leq & H(\lfloor\vect{X}\rfloor) + H\bigl(\lfloor \vect{X}+\eps\vect{Z}\rfloor \bigm| \lfloor\vect{X}\rfloor\bigr) \nonumber\\
& \leq & H(\lfloor\vect{X}\rfloor) + H(\vect{V}_{\eps}) \label{eq:ach_5}
\end{IEEEeqnarray}
and
\begin{IEEEeqnarray}{lCl}
H(\lfloor \vect{X}+\eps\vect{Z}\rfloor) & \geq & H(\lfloor\vect{X}\rfloor) - H\bigl(\lfloor\vect{X}\rfloor\bigm| \lfloor\vect{X}+\eps\vect{Z}\rfloor\bigr) \nonumber\\
& \geq & H(\lfloor\vect{X}\rfloor) - H(\vect{V}_{\eps}) \label{eq:ach_5.5}
\end{IEEEeqnarray}
where we define $\vect{V}_{\eps} \triangleq \lfloor \vect{X}+\eps\vect{Z}\rfloor - \lfloor\vect{X}\rfloor$. Note that $\vect{V}_{\eps}$ can also be written as $\vect{V}_{\eps}=\lfloor\bar{\vect{X}}+\eps\vect{Z}\rfloor$, where $\bar{\vect{X}} \triangleq \vect{X} -\lfloor\vect{X}\rfloor$.

In view of \eqref{eq:ach_5} and \eqref{eq:ach_5.5}, Part~(ii) of Lemma~\ref{lemma:cont} follows by showing that $H(\vect{V}_{\eps})$ vanishes as $\eps$ tends to zero. We begin by writing this entropy as (see, e.g., \cite[Eq.~(81)]{kostina15_arxiv})
\begin{equation}
\label{eq:app(ii)_0.25}
H(\vect{V}_{\eps}) = h\bigl(\lfloor\bar{\vect{X}}+\eps\vect{Z}\rfloor+\vect{U}\bigr)
\end{equation}
where $\vect{U}$ is a $d$-dimensional random vector that is uniformly distributed over the hypercube $[0,1)^d$ and that is independent of $(\vect{X},\vect{Z})$. We next show that
\begin{equation}
\label{eq:app(ii)_0.5}
\lim_{\eps\downarrow 0} h\bigl(\lfloor\bar{\vect{X}}+\eps\vect{Z}\rfloor+\vect{U}\bigr) = h(\vect{U}).
\end{equation}
The differential entropy of $\vect{U}$ is zero, so \eqref{eq:app(ii)_0.25} and \eqref{eq:app(ii)_0.5} demonstrate that $H(\vect{V}_{\eps})$ vanishes as $\eps$ tends to zero, which in turn proves Part~(ii) of Lemma~\ref{lemma:cont}.

Since conditioning reduces entropy \cite[Sec.~9.6]{coverthomas91}, we have
\begin{equation}
h\bigl(\lfloor\bar{\vect{X}}+\eps\vect{Z}\rfloor+\vect{U}\bigr) \geq h(\vect{U}).
\end{equation}
To prove \eqref{eq:app(ii)_0.5}, it thus remains to show that
\begin{equation}
\label{eq:app(ii)_CLAIM}
\varlimsup_{\eps\downarrow 0} h\bigl(\lfloor\bar{\vect{X}}+\eps\vect{Z}\rfloor+\vect{U}\bigr) \leq h(\vect{U}).
\end{equation}
To this end, we follow along the lines of the proof of Theorem~1 in \cite{linderzamir94} (see also the proof of Lemma~6.9 in \cite{lapidothmoser03_3}). Let the random vectors $\tilde{\vect{Y}}_{\eps}$ and $\tilde{\vect{Y}}_0$ have the respective pdfs
\begin{subequations}
\begin{IEEEeqnarray}{lCll}
f_{\tilde{\vect{Y}}_{\eps}}(\vect{y}) & = & \left(\frac{r}{d}\right)^{\frac{d}{r}-1}\frac{1}{V_d \Gamma(d/r) \sigma_{\eps}^{\frac{d}{r}}} e^{-\frac{d}{r\sigma_{\eps}}\|\vect{y}\|^r}, \quad & \vect{y}\in\Reals^d \\
f_{\tilde{\vect{Y}}_{0}}(\vect{y}) & = & \left(\frac{r}{d}\right)^{\frac{d}{r}-1}\frac{1}{V_d \Gamma(d/r) \E{\|\vect{U}\|^r}^{\frac{d}{r}}} e^{-\frac{d}{r\E{\|\vect{U}\|^r}}\|\vect{y}\|^r}, \quad & \vect{y}\in\Reals^d
\end{IEEEeqnarray}
\end{subequations}
where
\begin{equation}
\sigma_{\eps} \triangleq \E{\|\lfloor\bar{\vect{X}}+\eps\vect{Z}\rfloor + \vect{U}\|^r}.
\end{equation}
It follows that
\begin{equation}
\label{eq:app(ii)_D1}
D\bigl(f_{\lfloor\bar{\vect{X}}+\eps\vect{Z}\rfloor+\vect{U}}\bigm\| f_{\tilde{\vect{Y}}_{\eps}}\bigr) = \frac{d}{r} +\log\left(\frac{V_d\Gamma(d/r)}{(r/d)^{d/r-1}}\right) + \frac{d}{r}\log \sigma_{\eps} - h(\lfloor\bar{\vect{X}}+\eps\vect{Z}\rfloor+\vect{U})
\end{equation}
and
\begin{equation}
\label{eq:app(ii)_D2}
D\bigl(f_{\vect{U}}\bigm\| f_{\tilde{\vect{Y}}_{0}}\bigr) = \frac{d}{r} +\log\left(\frac{V_d\Gamma(d/r)}{(r/d)^{d/r-1}}\right) + \frac{d}{r}\log \E{\|\vect{U}\|^r} - h(\vect{U}).
\end{equation}
As we shall argue next, the pdf of $\lfloor\bar{\vect{X}}+\eps\vect{Z}\rfloor+\vect{U}$ converges pointwise to the pdf of $\vect{U}$ as $\eps$ tends to zero, so by Scheffe's lemma $\lfloor\bar{\vect{X}}+\eps\vect{Z}\rfloor+\vect{U} \Rightarrow\vect{U}$ as $\eps$ tends to zero. Indeed, the pdf of $\lfloor\bar{\vect{X}}+\eps\vect{Z}\rfloor+\vect{U}$ is given by
\begin{equation}
\label{eq:app(ii)_1}
f_{\lfloor\bar{\vect{X}}+\eps\vect{Z}\rfloor+\vect{U}}(\vect{x}) = \sum_{\vect{i}\in\Integers^d} \Prob\bigl(\lfloor\bar{\vect{X}}+\eps\vect{Z}\rfloor=\vect{i}\bigr)\I{\lfloor\vect{x}\rfloor=\vect{i}}, \quad \vect{x}\in\Reals^d.
\end{equation}
Since $\E{\|\vect{Z}\|^r}<\infty$, we have that $\eps\vect{Z}\to \vect{0}$ almost surely as $\eps$ tends to zero, which implies that $\eps\vect{Z}\Rightarrow \vect{0}$ as $\eps$ tends to zero. Furthermore, the independence of $\vect{X}$ and $\vect{Z}$ implies that $\bar{\vect{X}}+\eps\vect{Z} \Rightarrow \bar{\vect{X}}$ as $\eps$ tends to zero. Since by assumption $\Prob(\vect{X}\in\Integers^d)=0$, it follows that the probability $\Prob(\bar{\vect{X}}\in\Integers^d)$ is zero, so \cite[Th.~2.8.1, p.~122]{AsDo00}
\begin{equation}
\label{eq:app(ii)_2}
\lim_{\eps\downarrow 0} \Prob\bigl(\lfloor\bar{\vect{X}}+\eps\vect{Z}\rfloor=\vect{i}\bigr) = \Prob\bigl(\lfloor\bar{\vect{X}}\rfloor=\vect{i}\bigr) = \I{\vect{i}=\vect{0}}
\end{equation}
where the last step follows because, by definition, $\lfloor\bar{\vect{X}}\rfloor = \vect{0}$ almost surely. Applying \eqref{eq:app(ii)_2} to \eqref{eq:app(ii)_1}, and noting that $f_{\vect{U}}(\vect{u})=\I{\lfloor\vect{u}\rfloor=\vect{0}}$, $\vect{u}\in\Reals^d$, the claim that $f_{\lfloor\bar{\vect{X}}+\eps\vect{Z}\rfloor+\vect{U}}$ converges pointwise to $f_{\vect{U}}$ as $\eps$ tends to zero follows.

We next show that
\begin{equation}
\label{eq:app(ii)_2.5}
\lim_{\eps\downarrow 0} \E{\|\lfloor\bar{\vect{X}}+\eps\vect{Z}\rfloor + \vect{U}\|^r} = \E{\|\vect{U}\|^r}.
\end{equation}
To this end, we first note that, by the continuity of norms, and because the function $x\mapsto \lfloor x \rfloor$ is continuous for $x\notin \Integers$,
\begin{equation}
\label{eq:app(ii)_3} 
\lim_{\eps\downarrow 0} \|\lfloor \bar{\vect{x}}+\eps\vect{z}\rfloor+\vect{u}\|^r = \|\lfloor\bar{\vect{x}}\rfloor +\vect{u}\|^r = \|\vect{u}\|^r
\end{equation}
for every $\vect{z}\in\Reals^d$, $\vect{u}\in[0,1)^d$, and for $\bar{\vect{x}}\in(0,1)^d$. Furthermore, since on a finite-dimensional vector space any two norms are within a constant factor of one another \cite[p.~273]{hornjohnson91}, we have
\begin{equation}
\label{eq:app(ii)_4}
\underline{c}\, \|\lfloor \bar{\vect{x}}+\eps\vect{z}\rfloor+\vect{u}\|_1 \leq \|\lfloor \bar{\vect{x}}+\eps\vect{z}\rfloor+\vect{u}\| \leq \bar{c}\, \|\lfloor \bar{\vect{x}}+\eps\vect{z}\rfloor+\vect{u}\|_1
\end{equation}
for some constants $\bar{c}\geq\underline{c}> 0$, where $\|\vect{z}\|_1\triangleq |z_1|+\ldots + |z_d|$, $\vect{z}=(z_1,\ldots,z_d)\in\Reals^d$ denotes the $L_1$-norm. It thus follows that, for every $0<\eps\leq 1$,
\begin{IEEEeqnarray}{lCl}
\|\lfloor \bar{\vect{x}}+\eps\vect{z}\rfloor+\vect{u}\|^r & \leq & \bar{c}^r \|\lfloor \bar{\vect{x}}+\eps\vect{z}\rfloor+\vect{u}\|_1^r \nonumber\\
& \leq & \bar{c}^r \| |\vect{z}|+\vect{3}\|_1^r \nonumber\\
& \leq & \bar{c}^r \bigl(\|\vect{z}\|_1+\|\vect{3}\|_1\bigr)^r \nonumber\\
& \leq & \frac{\bar{c}^r}{\underline{c}^r} \bigl(\|\vect{z}\|+\|\vect{3}\|\bigr)^r \label{eq:app(ii)_5}
\end{IEEEeqnarray}
where $\vect{3}$ denotes the $d$-dimensional vector $(3,\ldots,3)$. Here the first step follows from \eqref{eq:app(ii)_4}; the second step follows because $|\lfloor x\rfloor|\leq |x|+1$, $x\in\Reals$ and because every component of $\bar{\vect{x}}$ and $\vect{u}$ satisfies $0\leq \bar{x}_{\ell},u_{\ell}< 1$; the third step follows from the triangle inequality; and the last step follows again from \eqref{eq:app(ii)_4}.

The lemma's assumptions $\E{\|\vect{Z}\|^r}<\infty$ and $\Prob(\vect{X}\in\Integers^d)=0$ imply that
\begin{equation}
\E{\bigl(\|\vect{Z}\|+\|\vect{3}\|\bigr)^r} < \infty
\end{equation}
and
\begin{equation}
\Prob\bigl(\bar{\vect{X}}\in(0,1)^d\bigr) = 1.
\end{equation}
Consequently, \eqref{eq:app(ii)_2.5} follows from \eqref{eq:app(ii)_3} and the dominated convergence theorem \cite[Th.~1.6.9, p.~50]{AsDo00}:
\begin{equation}
\lim_{\eps\downarrow 0} \E{\|\lfloor\bar{\vect{X}}+\eps\vect{Z}\rfloor + \vect{U}\|^r} = \E{\lim_{\eps\downarrow 0} \|\lfloor\bar{\vect{X}}+\eps\vect{Z}\rfloor + \vect{U}\|^r} = \E{\|\vect{U}\|^r}.
\end{equation}
Since $f_{\tilde{\vect{Y}}_{\eps}}$ is a continuous function of $\sigma_{\eps}$, \eqref{eq:app(ii)_2.5} implies that $f_{\tilde{\vect{Y}}_{\eps}}$ converges pointwise to $f_{\tilde{\vect{Y}}_{0}}$ as $\eps$ tends to zero, so by Scheffe's lemma $\tilde{\vect{Y}}_{\eps}\Rightarrow \tilde{\vect{Y}}_{0}$ as $\eps$ tends to zero.

We conclude that $\lfloor\bar{\vect{X}}+\eps\vect{Z}\rfloor+\vect{U} \Rightarrow\vect{U}$ and $\tilde{\vect{Y}}_{\eps}\Rightarrow \tilde{\vect{Y}}_{0}$ as $\eps$ tends to zero, so the lower semicontinuity of relative entropy gives
\begin{equation}
\label{eq:app(ii)_6}
\varliminf_{\eps\downarrow 0} D\bigl(f_{\lfloor\bar{\vect{X}}+\eps\vect{Z}\rfloor+\vect{U}}\bigm\| f_{\tilde{\vect{Y}}_{\eps}}\bigr) \geq D\bigl(f_{\vect{U}}\bigm\| f_{\tilde{\vect{Y}}_{0}}\bigr).
\end{equation}
Combining \eqref{eq:app(ii)_6} with \eqref{eq:app(ii)_D1} and \eqref{eq:app(ii)_D2}, and using that, by \eqref{eq:app(ii)_2.5}, $\sigma_{\eps} \to \E{\|\vect{U}\|^r}$ as $\eps$ tends to zero, it follows that
\begin{equation}
\varlimsup_{\eps\downarrow 0} h(\lfloor\bar{\vect{X}}+\eps\vect{Z}\rfloor+\vect{U}) \leq h(\vect{U}).
\end{equation}
This proves Part~(ii) of Lemma~\ref{lemma:cont}.

\section*{Acknowledgment}
The author wishes to thank Helmut B\"olcskei, David Stotz, and Gonzalo Vazquez-Vilar for helpful discussions. The author further wishes to thank  Giuseppe Durisi and Tam\'as Linder for calling his attention to references \cite{wuverdu10_2} and \cite{csiszar61}, respectively.

\bibliographystyle{IEEEtran}         % in order of first citation
\bibliography{/Users/koch/Library/texmf/tex/bibtex/header_short,/Users/koch/Library/texmf/tex/bibtex/bibliofile}

\end{document}